# Tunable multiband directional electromagnetic scattering from spoof Mie resonant structure

Hong-Wei Wu,[1,3*] Hua-Jun Chen,[1] Hua-Feng Xu,[1] Yu Zhou,[2,3] Dong-Xiang Qi[4] and Yue-Ke Wang[4]

[1]*School of Mechanics and Photoelectric Physics, Anhui University of Science and Technology, Huainan 232001, China*
[2]*School of Science, Hangzhou Dianzi University, Hangzhou 310018, China*
[3]*National Laboratory of Solid State Microstructures and School of Physics, Collaborative Innovation Center of Advanced Microstructures, Nanjing University, Nanjing 210093, China*
[4]*School of Science, Jiangnan University, Wuxi 214122, China*

we demonstrate that directional electromagnetic scattering can be realized from a artificial Mie resonant strcuture which supports electric and magnetic dipole modes simultaneously. The directivity of the far-field radiation pattern can be switched by changing the incident light wavelength as well as tailoring the geometric parameters of the structure. Particularly, the electric quadrupole at higher frequency contribute significantly to the scattered fields, leading to enhancement of the directionality. In addition, we further design a quasiperiodic spoof Mie resonant structure by alternately inserting two materials into the slits. The results show that multi-band directional light scattering are realized by exciting multiple electric and magnetic dipole modes with different frequencies in the quasiperiodic structure. The presented design concept is general from microwave to terahertz region and can be applied for various advanced optical devices, such as antenna, metamaterial and metsurface.

[*]Corresponding author: hwwu@aust.edu.cn

## I. INTRODUCTION

Directional electromagnetic scattering, which can be induced by interference between the electric and magnetic dipolar resonances, plays a key role in many fundamental and applied researches [1, 2]. In 1983, hypothetical magneticdielectric particle which exhibiting electric and magnetic dipolar resonances had been theoretically proposed by Kerker *et al* to predict the unidirectional forward or backward scattering due to the constructive interference of resonances enhancing scattering in one direction and destructive interference minimizing the scattered intensity in the opposite direction [3]. In order to obtain the directional light scattering, various metamaterials based on metallic nanostructures have been engineered to exhibit artificial magnetism along with their intrinsic electrical response in the visible and infrared regime [4-7]. However, a main drawback of using plasmonic nanostructures is the intrinsic losses, which strongly limits their practical use in directional light scattering. Fortunately, using high-index dielectric nanoparticles can avoid such limitation and exhibit both electric and magnetic dipole resonances in same particle simultaneously [8-12]. This characteristic makes high-index dielectric nanoparticles as popular choices for realizing the directional light scattering in visible frequencies. Recently, various high-index dielectric nanostructures have been theoretical and experimental investigated for directional light scattering, such as nanodisk [13], nanowire [14], nanosphere [15]and sphere/disk dimers [16-18].

Moreover, in order to transfer the exotic features of conventional localized surface plasmons (LSPs) to microwave and terahertz regions, Pors *et al* have proposed the concept of spoof LSPs based on the textured perfect electric conductors (PECs) cylinder to mimic the metallic nanoparticles in the microwave regime [19]. Since then, extensive theoretical and experimental works have been performed to verify the existence of spoof LSPs in various textured metallic structures [20-24]. In addition, we also have demonstrated that a hollow metallic cylinder corrugated by periodic cut-through slits can supports the magnetic and electric dipole resonances similar to the subwavelength dielectric nanoparticles with a high-refractive index at optical frequencies, as the spoof Mie resonant structure [25]. Certainly, the optical properties for the high-index nanoparticles in the near-infrared or visible regimes can be rescaled to big-size particles in low frequencies with similar radius to wavelength ratios. For instance, a single low-loss dielectric subwavelength ceramic particle with diameter of 18mm and refractive index of 4 has been experimentally demonstrated to induce directional electromagnetic scattering in microwave regime [26]. However, the artificial Mie resonant structure has a greater degree of freedom for tuning resonance properties by changing the geometry parameters including radius, slits width and filling medium, compared with to natural high-index dielectric particles. Therefore, it may be significative to investigate the directional electromagnetic scattering from the spoof Mie resonant structure because the highly asymmetric scattering can be expediently tuned by adjusting the structure parameters.

In this paper, we demonstrate that electric and magnetic resonant modes can be simultaneously excited in spoof Mie resonant structure to induce directional electromagnetic scattering at low frequency. The results show that the unidirectional scattering can be tuned by tailoring the geometric parameters of the structure. Furthermore, the electric quadrupole at higher frequency contributes significantly to the forward scattered field, leading to the enhanced directionality of forward scattering comparing with one at lower frequency obtained by interference between pure electric and magnetic dipolar resonances. In order to realize multi-band directional electromagnetic scattering, we also propose a quasiperiodic spoof Mie resonant structure by alternately inserting different dielectric materials into slits which support multiple electric and magnetic dipolar resonance. When the water as dielectric material in the metallic slits, the forward or backward scattering from these structures can be active controlled by tuning the water temperature at microwave region because the refractive index of water is temperature-dependent in this region. Our results provide a potential approach for designing antenna of directional scattering in the microwave and terahertz region.

## II. ELECTROMAGNETIC RESPONES OF HOLLOW TEXTURED METALIC CYLINDER

We first consider a hollow 2D PEC cylinder with periodic cut-through slits as shown in Fig. 1(a). The hollow PEC cylinder with outer radius $R$ and inner radius $r$ is decorated with a set of radial slits with depth $h = R - r$, width $a$, and periodicity $d = 2\pi R / N$ (where $N$ are the total number of slits). A dielectric material with a refractive index of $n_s$ is filled into the slits, and the surround of this structure is air (i.e., both regions I and III are air). It is well known that the region II can also be regarded as a metamaterial of thickness $h = R - r$ with $\varepsilon_\rho = \infty$, $\varepsilon_\varphi = n_s^2 d / a$, and $\mu_z = a / d$ for the transverse magnetic (TM) polarized incident wave. ($\rho$, $\varphi$) are the polar coordinates as shown in Fig. 1(a). Without loss of generality, the structure parameters are chosen as $N = 30$, $r = 0.1R$, $n_s = 4$, and $a = 0.8d$ unless otherwise specified in this paper. The refractive index $n_s = 4$ of dielectric is chosen in slits in order to approach the one of silicon, then the structure can be simply seen as by inserting the metal strips into hollow silicon cylinder periodically. The outer radius $R$ is set to the unit length for generosity, and the frequency can be expressed in term of the dimensionless quantity $k_0 R$ where $k_0 = \omega / c$ ($\omega$ and $c$ are angular frequency and vacuum speed of light). The scale of whole structure are chosen in deep subwavelength scale. In order to probe the resonance mode, we here consider a TM-polarized incident plane wave propagated from top to bottom along $y$ direction. The scattering responses and field distributions of this structure are calculated by using the commercial software COMSOL MULTIPHYSICS based on finite element method (FEM).

The electromagnetic response of our structure with negligible absorptions can be accurately described by the scattering cross section (SCS). Figure 1(b) shows the normalized scattering cross section (SCS) as a function of the dimensionless quantity $k_0 R$ for the spoof Mie resonant structure with periodic slits. The SCS is normalized to the diameter $2R$ of the structure. We first give the total SCS (red solid line) of the structure based on numerical calculation in Fig. 1(b). It is obvious that three resonant peaks in the SCS spectrum correspond to the magnetic dipole mode, electric dipole mode and electric quadrupole mode from left to right. Furthermore, in order to gain a deeper insight into the resonant response of magnetic and electric modes, we present an effective medium theoretical model for the analytical description of our structure by employing the modal expansion technique [19]. In the limit of $a < d \ll \lambda_0$, the structure can be treated as a homogeneous and anisotropy metamaterial with $\varepsilon_\rho = \infty$, $\varepsilon_\varphi = n_s^2 \, d / a$, and $\mu_z = a / d$ in region II of Fig. 1(a) for the TM-polarized incident light. Inside regions I and III, Maxwell's equations can be decomposed into the free-space Helmholtz's equation for $H_z$ component. Solving the second order differential equations, the $H_z$ is given by considering the finite energy in region I and the Sommerfeld radiation conditions in region III, which are shown as:

$$H_z^I(\rho,\varphi) = \sum_{m=-\infty}^{\infty} A_m J_m(k_0 \rho)\exp(im\varphi), \tag{1}$$

$$H_z^{III}(\rho,\varphi) = \sum_{m=-\infty}^{\infty} D_m H_m^{(1)}(k_0 \rho)\exp(im\varphi), \tag{2}$$

where $J_m$ and $H_m^{(1)}$ are the Bessel and Hankel functions of first kind, respectively, and $A_m$ and $D_m$ are complex constants. $m$ is an integer quantifying the orbital angular momentum of the mode in the regions I and III, respectively. In the metamaterial of region II, the magnetic field can be simply written as [27]:

$$H_z^{II}(\rho,\varphi) = \sum_{m=-\infty}^{\infty} \left( B_m J_m(k_0 \rho) + C_m Y_m(k_0 \rho) \right)\exp(im\varphi), \tag{3}$$

where $Y_m$ is the Bessel function of the second kind, $B_m$ and $C_m$ are complex constants. After matching boundary conditions on $H_z$ and $E_\varphi$ at inner and outer boundaries ($\rho = r$, $\rho = R$), we can give the analytic expression of the SCS as:

$$\sigma_{sc} = \frac{4}{k_0} \sum_{m=-\infty}^{\infty} |D_m|^2, \tag{4}$$

the complex constant $D_m$ can be writed as:

$$D_m = -i^m \frac{\dfrac{a}{d} p J_m(k_0 R) + q n_s J_m^{'}(k_0 R)}{\dfrac{a}{d} p H_m^{(1)}(k_0 R) + q n_s H_m^{(1)'}(k_0 R)}. \tag{5}$$

where $p = J_1(k_0 n_s R) - M Y_1(k_0 n_s R)$, $q = J_0(k_0 n_s R) - M Y_0(k_0 n_s R)$ and M = [J'$_m$(k$_0$r) J$_0$(k$_0$n$_s$r) / J$_m$(k$_0$r) + a J$_1$(k$_0$n$_s$r) / (n$_s$d)] / [J'$_m$(k$_0$r) Y$_0$(k$_0$n$_s$r) / J$_m$(k$_0$r) + a Y$_1$(k$_0$n$_s$r) / (n$_s$d)]. $J_0$ and $J_1$ ($Y_0$ and $Y_1$) being the zero- and first- order Bessel functions of first (second) kind, respectively. In Fig. 1(b), the analytical total SCS is shown with black solid curve calculated from Eq. (4). Compared with red solid line and black solid line, we can find that they exhibit the similar spectra features, except the magnitude of peak corresponding to magnetic dipole mode. The difference between simulative SCS and analytical SCS at magnetic dipole resonance may be from the selected value of $\varepsilon_\rho$ and boundary settings in simulation. It is well known that $m = 0$, $m = 1$, and $m = 2$ in Eq. (4) respectively correspond to magnetic dipole, electric dipole and electric quadripole [23], then three peaks in analytical SCS can be identified by calculating the SCS of individual resonance mode as shown in blue dash, green dash, and pink dash lines in Fig. 1(b), respectively.

## III. DIRECTIONAL SCATTERING INDUCED FROM HOLLOW TEXTURED METALIC CYLINDER

### A. Directional scattering induced from periodic structure

In the following, we will discuss the directional scattering from the designed structure based on these resonant modes demonstrated in Fig. 1(b). Figure 2(a) shows scattering properties of the structure in free space calculated by numerical simulation. The structure is excited by a plane wave from top and the scattering into the upper (backward) or lower (forward) hemispheres are calculated as shown in the inset. In Fig. 2(a), the blue and red solid lines indicate the forward and backward scattering cross sections respectively. The black solid line is the forward-to-backward (F/B) ratio. Then, we can find that there are three well defined spectral ranges with different scattering properties. Firstly, in the spectral range with $k_0R < 0.56\pi/3$, the far-field scattering is dominated by the forward scattering. Particularly, the backward scattering is almost zero and the F/B ratio is maximum value at $k_0R = 0.48\pi/3$. In order to verify this conclusion, we calculate the angular scattering diagram in Fig. 2(b) denoted as "1" corresponding to that in Fig. 2(a). It is obvious that the scattering light is completely transformed into forward scattering and barely no backward scattering. At the location $k_0R = 0.56\pi/3$ denoted by vertical dotted line "2", the forward scattering equals to backward scattering and F/B = 1. We can confirm that this location corresponds to the magnetic dipolar resonance from Fig. 1(b) and the far-field pattern denoted as "2" in Fig. 2(b). Next, in the second spectral range $0.56\pi/3 < k_0R < 0.696\pi/3$, the backward scattering is dominant and the F/B < 1. It is not difficult to see that the minimum value of F/B ratio can be obtained for $k_0R = 0.62\pi/3$. From the far field scattering denoted as "3" in Fig. 2(b), we can see that the backward scattering is significantly enhanced while the forward scattering is considerably suppressed. The physical mechanism behind the directional scattering corresponding to $k_0R = 0.48\pi/3$ and $0.62\pi/3$ in the first and second spectral range is the first and second Kerker condition described in Ref. [28] because the contribution of electric quadripole is a negligible quantity in these ranges. The symmetrical scattering in forward and backward direction occur at $k_0R = 0.696\pi/3$ corresponding to the electric dipolar resonance as can be seen in "4" of Fig. 2(b). However, in the third spectral range with $0.696\pi/3 < k_0R < 0.83\pi/3$, the maximum value of F/B ratio is greater than one of the first spectral range at $k_0R = 0.76\pi/3$, and the far field is also be polted in Fig. 2(b) denoted as "5". The result indicates that the scattering light from the spoof Mie resonant structure with this frequency is shaped into the forward direction and no backward scattering. While the reason of this zero-backward scattering is the mutual interference among the magnetic dipole, electric dipole and electric quadripole described as general Kerker condition ($p_x - m_z/c + ik_0Q_{xy}/6 = 0$, where $p_x$, $m_z$ and $Q_{xy}$ represent electric dipole moment, magnetic dipole moment and electric quadrupole moment, $c$ is the light speed) in Ref. [29], unlike the mechanism of the zero-backward scattering at $k_0R = 0.48\pi/3$. Though the F/B ratio starts to drop again for $k_0R > 0.83\pi/3$, this region is accompanied by a reduction of total scattering making it less attractive.

Until now, we have demonstrated that the directional electromagnetic scattering can be induced due to the simultaneous excitation and mutual interference of electric dipole, magnetic dipole and electric quadrupole. Next, we discuss the influence of structure parameters of the hollow textured PEC cylinder on the directional scattering. Figure 3(a) shows that forward and backward scattering cross section represented by blue and red solid line as the function of the dimensionless parameter $k_0R$ together with the F/B ratio curve for the structure with outside radius R. When decreasing the outside radius from R to 0.7R and keeping other parameters unchanged, we can find that three curves coincidentally blue-shift due to the increase of resonant frequencies corresponding to magnetic dipole, electric dipole and electric quadrupole as can be seen from Figs. 3(a) to 3(d). However, the phenomenons are completely different for increasing the inner radius as shown in Figs. 3(e)-3(h), the first spectral range redshifts and the third spectral range blueshifts with the increase of the second spectral width. In fact, the spectral shift mainly depends on the shift of these resonant modes after changing the structure parameters. It is well known that the resonant frequencies of the electric dipole and quadrupole are increased with decreasing the depth of the metallic slits [25] or grooves [19] whether by decreasing the outside radius or increasing the inner radius of the hollow textured PEC cylinder. While the resonant frequency of magnetic dipolar mode mainly depends on the radius of displacement current circle excited by electric field. Thus, the resonant frequency of magnetic dipolar mode increases with the decrease of outside radius of the structure because the radius of the displacement current circle is inwards compressed. Inversely, the radius of the displacement current circle is pushed outward with increasing the inner radius of this structure and leading to the redshift of magnetic dipolar mode as shown in Figs. 3(e)-3(h).

**B. Directional scattering induced from quasiperiodic structure**

Furthermore, we also design the quasiperiodic spoof Mie resonant structure by alternately inserting dielectric materials A and B into the slits as in Fig. 4(a). The quasiperiodic structure unit is described in detail in Fig. 4(b), the green and yellow regions represent the dielectric materials A and B. Figures 4(c) and 4(d) correspond to the perodic substructures solely filled dielectric A and B in $N/2$ slits, respectively. With our intuitive understanding, the quasiperiodic structure should possesses two sets of magnetic and electric dipolar resonant modes supported by slits A and B. In order to confirm this conjecture, we plot the SCSs of the quasiperiodic structure, perodic substructure A and B as black, red and blue solid line in Fig. 4(e) for setting dielectric materials $n_{sA} = 6$ and $n_{sB} = 4$, correspondingly. It is not difficult to find that the SCS of the quasiperiodic structure presents two magnetic dipolar modes and two electric dipolar modes with different frequencies. By comparing the three SCSs, we can also find that the electric dipolar modes of the quasiperiodic structure are consistent with those of the substructures, while all magnetic resonant modes are blue shifted. In fact, the shift of the magnetic resonant peaks of the quasiperiodic structure is the result of the contribution of additional displacement current induced by the other substructure filled with different material.

It has been demonstrated that the quasiperiodic structure can supports two sets of magnetic and electric dipolar resonances as above discussion. Naturally, the multiband directional scattering from the quasiperiodic structure can be expected by the mutual interference between the electric and magnetic dipolar modes. Figure 5(a) shows the forward and backward scattering cross section as the function of the dimensionless parameter $k_0R$ together with the F/B ratio curve. As we desired, there are four peaks (vertical green dashed lines $F_1$, $F_2$, $F_3$ and $F_4$) and three valleys (vertical orange dashed lines $B_1$, $B_2$ and $B_3$) in the F/B curve. This is easy to understand that the peaks $F_1$, $F_2$ and valley $B_1$ of the F/B curve are the result of interference between the magnetic and electric dipolar modes excited in slits A. Similarly, the peaks $F_3$, $F_4$ and valley $B_3$ come from the interference between magnetic and electric modes supported in slits B. However, the reason of forming valley $B_2$ is the interference between the electric dipolar mode supported by slits A and the magnetic dipolar mode supported by slits B. In order to verify the performance of directional scattering, we plot the far field pattern corresponding to the peaks and valleys of the F/B curve in Fig. 5(b). The calculated angular scattering diagrams at the locations $F_1$, $F_2$, $F_3$ and $F_4$ consistently demonstrate almost perfect alignment of the scattering into the forward direction. The scattering diagrams at the locations $B_1$, $B_2$ and $B_3$ demonstrate the scattering alignment into the backward direction with minor residual scattering in the forward direction.

Water is characterized by a refractive index at a temperature below 100°C in microwave band [30, 31], such as refractive index 7.07 at 90°C and 8.9 at 20°C in the frequency range from 1GHz to 6GHz. In order to active control the directional scattering by tuning water temperature, we design the outside radius R of the hollow textured strcuture as 5 millimeter, and fill the metallic slits with the water in the perodic structure as indicated in Fig. 1(a). We calculate the F/B curvers as the function of frequency for different refractive indexs ($n_s$ = 7, 7.5, 8) of the warter by controlling water temperatures from 90°C to 25°C. The result shows the whole F/B curve redshift due to the increase of refractive index of water as can be seen in Fig. 6(a). In the quasiperiodic structure, we choose silicon as the dielectric material A and water as the dielectric material B. By changing the temperature of the water, we can find that the peaks and valleys in the F/B curve are red shifted except the fourth peak for increasing the refractive index of water from $n_{sB}$ = 7 to $n_{sB}$ = 8 in Fig. 6(b). The redshift of the first two peaks is the result of the redshift of magnetic and electric resonant modes for increasing the refractive index of water, while the redshift of the third peak comes from the influence of slits B on the magnetic dipolar resonance supported in slits A as shown in Fig. 4(e), even if the refractive index of silicon is not affected by temperature.

## IV. CONCLUSION

In conclusion, we have demonstrated that the directional scattering can be induced by the spoof Mie resonant structure. The directional scattering behavious results from interference of electric and magnetic dipoles supported by the designer structure. The simulation results confirm the theoretical predicitions, which state that the structure can scatter the most of light in the forward and backward direction at different frequencies. The directionality of the anisotropic scattering can be conveniently controlled by tailoring the geometric and material parameters. Furthermore, we also propose a quasiperiodic structure to present multiband directional scattering with the excitation of multiple electric and magnetic dipolar resonances. For some practical application, water whose refractive index depended on the temperature in microwave region is filled into the slits for realizing the active control function. These results may open a novel route towards manipulation and control of electromagnetic scattering in microwave and terahertz region.

**ACKNOWLEDGMENTS**

This work was supported by the National Natural Science Foundation of China (Grant Nos. 11647117, 11747065 and 11647001), and the Natural Science Foundation of Anhui Province (Nos. 1708085QA11 and 1508085QF140).

1. A. I. Kuznetsov, A. E. Miroshnichenko, M. L. Brongersma, Y. S. Kivshar, and B. Luk'yanchuk, Science **354**, aag2472, (2016).

2. Z. J. Yang, R. B. Jiang, X. L. Zhuo, Y. M. Xie, J. F. Wang, and H. Q. Lin, Phys. Rep. **701**, 1 (2017).

3. M. Kerker, D. S. Wang, and C. L. Giles, J. Opt. Soc. Am. **73**, 765 (1983).

4. W. Liu, J. Zhang, B. Lei, H. Ma, W. Xie, and H. Hu, Opt. Express **22**, 16178 (2014).

5. W. Liu, A. E. Miroshnichenko, D. N. Neshev, and Y. S. Kivshar, ACS Nano **6**, 5489 (2012).

6. R. Gomez-Medina, B. Garcia-Camara, I. Suarez-Lacalle, F. Gonzalez, F. Moreno, M. Nieto-Vesperinas, and J. J. Saenz, J. Nanophotonics **5**, 053512 (2011).

7. A. Alu and N. Engheta, Opt. Express **17**, 5723 (2009).

8. Q. Zhao, J. Zhou, F. Zhang, and D. Lippens, Mater. Today **12**, 60 (2009).

9. C. M. Soukoulis and M. Wegener, Nat. Photonics **5**, 523 (2011).

10. A. García-Etxarri, R. Gómez-Medina, L. S. Froufe-Pérez, C. López, L. Chantada, F. Scheffold, J. Aizpurua, M. Nieto-Vesperinas, and J. J. Sáenz, Opt. Express **19**, 4815 (2011).

11. J. C. Ginn, I. Brener, D. W. Peters, J. R. Wendt, J. O. Stevens, P. F. Hines, L. I. Basilio, L. K. Warne, J. F. Ihlefeld, P. G. Clem, and M. B. Sinclair, Phys. Rev. Lett. **108**, 097402 (2012).

12. C. Wang, Z. Y. Jia, K. Zhang, Y. Zhou, R. H. Fan, X. Xiong, and R. W. Peng, J. Appl. Phys. **115**, 244312 (2014).

13. I. Staude, A. E. Miroshnichenko, M. Decker, N. T. Fofang, S. Liu, E. Gonzales, J. Dominguez, T. S. Luk, D. N. Neshev, I. Brener, and Y. Kivshar, ACS Nano **7,** 7824 (2013).

14. W. Liu, Phys. Rev. A **96,** 023854 (2017).

15. Y. H. Fu, A. I. Kuznetsov, A. E. Miroshnichenko, Y. F. Yu, and B. Luk'yanchuk, Nat. Commun. **4**, 1527 (2013).

16. J. H. Yan, P. Liu, Z. Y. Lin, H. Wang, H. J. Chen, C. X. Wang, and G. W. Yang, ACS Nano **9** 2968 (2015).

17. B. García-Cámara, J. F. Algorri, A. Cuadrado, V. Urruchi, J. M. Sánchez-Pena, R. Serna, and R. Vergaz, J. Phys. Chem. C **119**, 19558 (2015).

18. T. Shibanuma, P. Albella, and S. A. Maier, Nanoscale **8**, 14184 (2016).

19. A. Pors, E. Moreno, L. Martin-Moreno, J. B. Pendry, and F. J. Garcia-Vidal, Phys. Rev. Lett. **108**, 223905 (2012).

20. X. P. Shen and T. J. Cui, Laser Photon. Rev. **8**, 137 (2014).

21. Z. Gao, F. Gao, H. Xu, Y. Zhang, and B. Zhang, Opt. Lett. **41**, 2181 (2016).

22. Z. Li, B. Xu, L. Liu, J. Xu, C. Chen, C. Gu, and Y. Zhou, Sci. Rep. **6**, 27158 (2016).

23. P. A. Huidobro, X. P. Shen, J. Cuerda, E. Moreno, L. Martin-Moreno, F. J. Garcia-Vidal, T. J. Cui, and J. B. Pendry, Phys. Rev. X **4**, 021003 (2014).

24. H. W. Wu, H. J. Chen, H. Y. Fan, Y. Li, and X. W. Fang, Opt. Lett. **42**, 791 (2017).

25. H. W. Wu, Y. Z. Han, H. J. Chen, Y. Zhou, X. C. Li, J. Gao, and Z. Q. Sheng, Opt. Lett. **42**, 4521 (2017).

26. J. M. Geffrin, B. García-Cámara, R. Gómez-Medina, P. Albella, L. S. Froufe-Pérez, C. Eyraud, A. Litman, R. Vaillon, F. González, M. Nieto-Vesperinas, J. J. Sáenz, and F. Moreno, Nat. Commun. **3**, 1171 (2012).

27. H. W. Wu, F. Wang, Y. Q. Dong, F. Z. Shu, K. Zhang, R. W. Peng, X. Xiang, and M. Wang, Opt. Express **23**, 32087 (2015).
28. M. Kerker, D. S. Wang, and C. L. Giles, J. Opt. Soc. Am. **73,** 765 (1983).
29. R. Alaee, R. Filter, D. Lehr, F. Lederer, and C. Rockstuhl, Opt. Lett. **40**, 2645 (2015).
30. M. V. Rybin, D. S. Filonov, P. A. Belov, Y. S. Kivshar, and M. F. Limonov, Sci. Rep. **5**, 8774 (2015).
31. A. Andryieuski, S. M. Kuznetsova, S. V Zhukovsky, Y. S Kivshar, and A. V Lavrinenko, Sci. Rep. **5**, 13535 (2015).

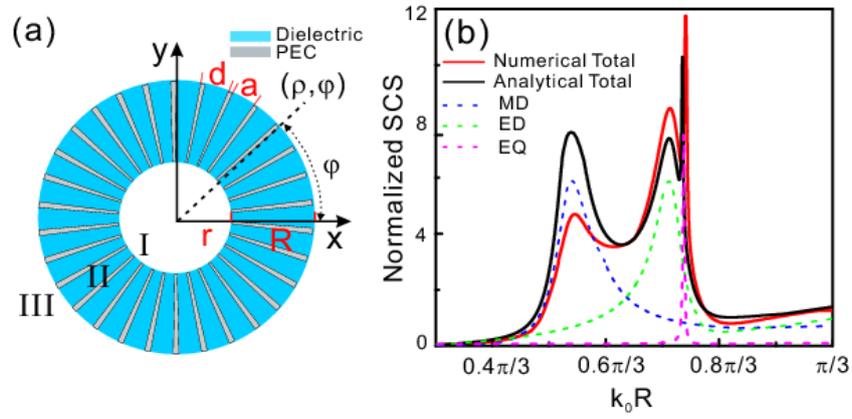

Fig. 1. (a) A two-dimensional corrugated PEC hollow cylinder with periodic cut-through slits with the inner and outer radii $r$ and $R$, periodicity $d$, and slit width $a$. Blue and gray regions represent the dielectric and PEC materials. (b) The calculated SCS spectrum for the textured PEC hollow cylinder. Red solid line corresponds to simulation result of textured PEC hollow cylinder and black solid line is the analytic calculation for the structure parameters $N = 30$, $r = 0.1R$ and $a = 0.8d$. The refractive index of dielectric material in slits is $n_s = 4$. Blue dash line, green dash line and pink dash line correspond to the SCS of modes m = 0, m = 1 and m =2, respectively.

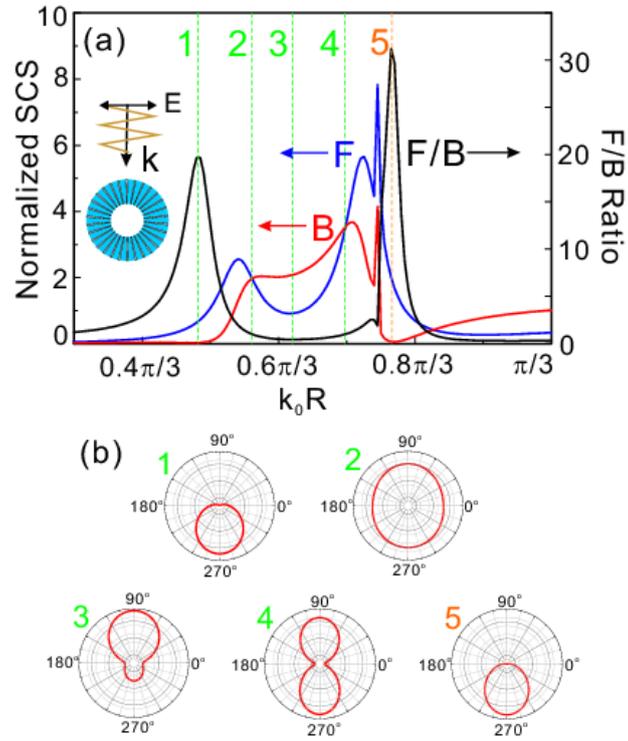

Fig. 2. (a) Forward (blue curve) and the backward (red curve) scattering cross-sections, and the forward-to-backward ratio (black curve) of the spoof localized plasmonic structure. The inset in (a) indicates that a TM-polarized incident plane wave propagated from top to bottom along *y* direction. (b) Far-field scattering patterns in the five spectral points corresponding to those vertical dashed lines in (a) denoted by "1", "2", "3", "4" and "5".

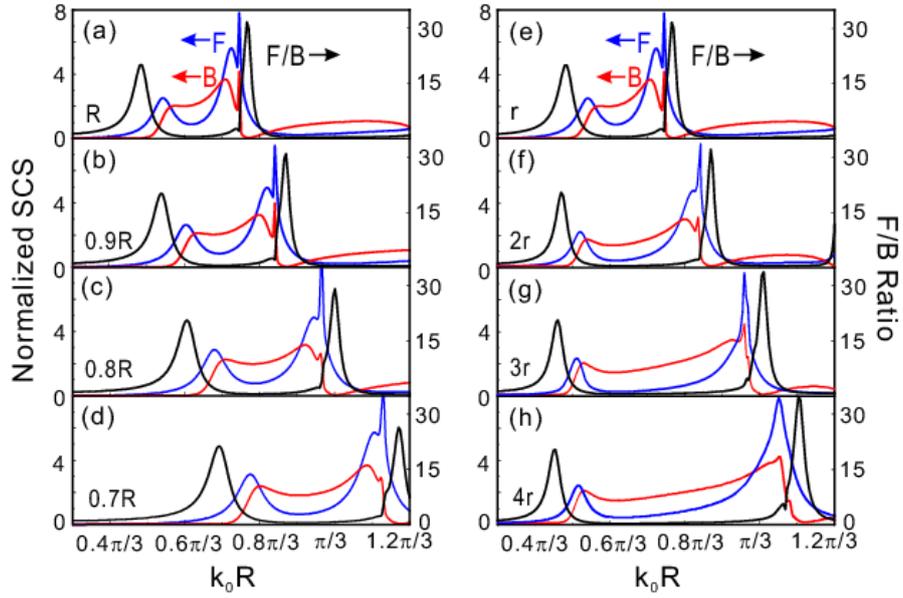

Fig. 3. Forward (blue curve) and backward (red curve) scattering cross-sections as a function of dimensionless quantity $k_0R$ together with the forward-to-backward ratio for the designer structures with outside radius R, 0.9R, 0.8R and 0.7R in (a)-(d), with different inner radius r, 2r, 3r and 4r in (e)-(f). While other parameters remain the same.

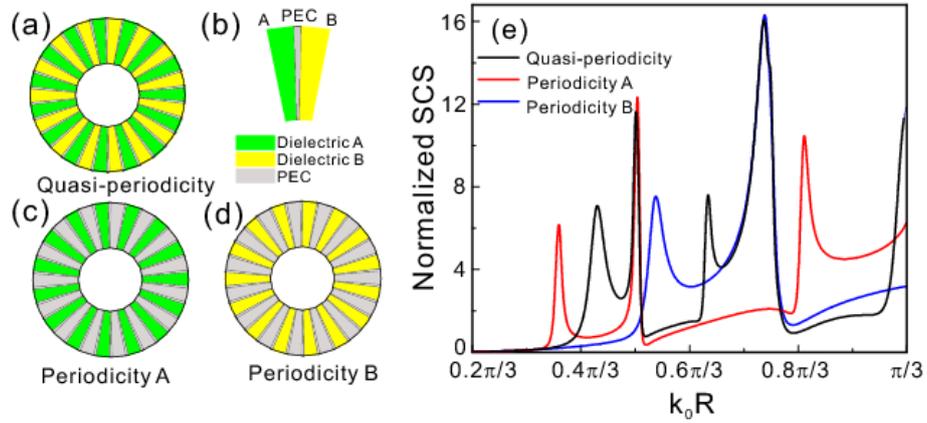

Fig. 4. (a) Schematic diagram of quasiperiodic structure alternately filled with dielectric A and B in slits. The quasiperiodic structure unit is detailedly described in (b), the green and yellow regions represent the dielectric materials A and B. (c) and (d) Two periodic substructures with one slit in one periodicty filled with dielectric A and B, correspondingly. The calculated SCS of the quasiperiodic structure, periodic substructure A and B as a function of dimensionless quantity $k_0R$.

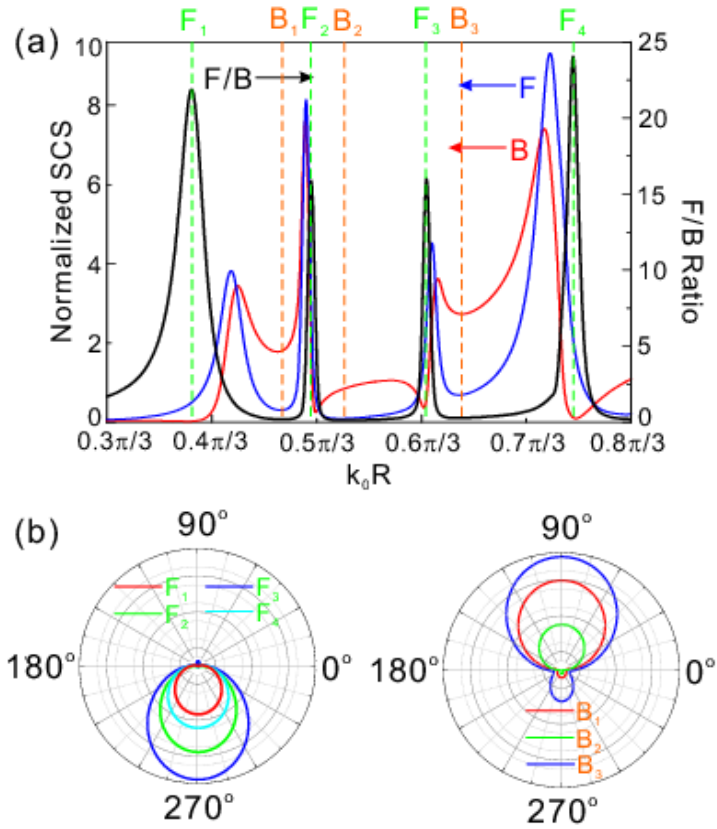

Fig. 5. (a) Forward (blue curve) and the backward (red curve) scattering cross-sections, and the forward-to-backward ratio (black curve) of the quasiperiodic spoof localized plasmonic structure. (b) Far-field scattering patterns in these spectral points corresponding to green dashed lines in (a) denoted by "$F_1$", "$F_2$", "$F_3$", "$F_4$" and orange dashed lines "$B_1$", "$B_2$", "$B_3$".

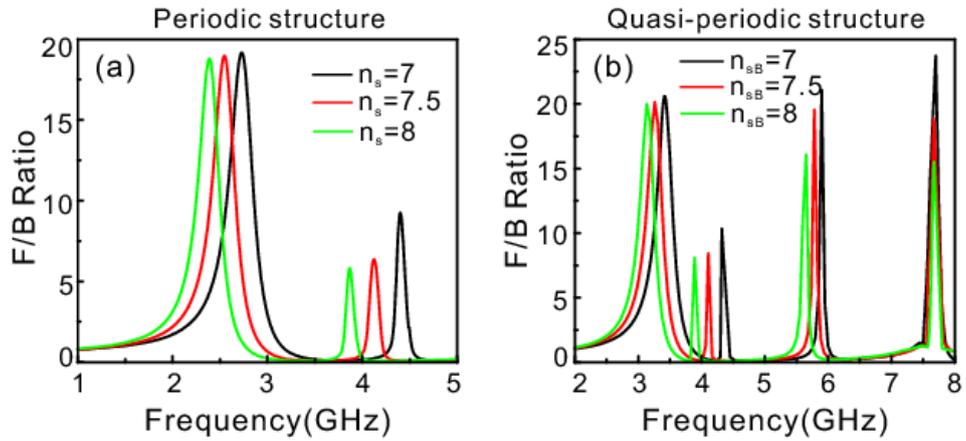

Fig. 6. (a) F/B curvers as the function of frequency for different refractive indexs ($n_s$ = 7, 7.5, 8) of the warter filled in the slits of the periodic structure as shown in Fig. 1(a). (b) F/B curvers as the function of frequency for different refractive indexs ($n_{sB}$ = 7, 7.5, 8) of the warter as the dielectric material B and silicon as the dielectric material A in the quasiperiodic structure indicated in Fig. 4(a).